\documentclass[fleqn,usenatbib]{mnras}

\usepackage{newtxtext,newtxmath}
\usepackage{mathrsfs}

\usepackage[T1]{fontenc}
\usepackage{ae,aecompl}
\usepackage{booktabs}

\usepackage{graphicx}	
\usepackage{amsmath}	
\usepackage{amssymb}	

\title[Luminosity-duration relation of fast radio bursts]{Luminosity-duration relation of fast radio bursts}

\author[T. Hashimoto et al.]{
Tetsuya Hashimoto,$^{1}$\thanks{E-mail: tetsuya@phys.nthu.edu.tw}
Tomotsugu Goto,$^{1}$
Ting-Wen Wang,$^{1}$
Seong Jin Kim,$^{1}$
\newauthor
Yi-Han Wu$^{1}$
and Chien-Chang Ho$^{1}$
\\
$^{1}$Institute of Astronomy, National Tsing Hua University, 101, Section 2. Kuang-Fu Road, Hsinchu, 30013, Taiwan (R.O.C.)\\
}

\date{Accepted 2019 June 17. Received 2019 June 04; in original form 2019 March 8}

\pubyear{2019}

\begin{document}
\label{firstpage}
\pagerange{\pageref{firstpage}--\pageref{lastpage}}
\maketitle

\begin{abstract}
Nature of dark energy remains unknown.
Especially, to constrain the time variability of the dark-energy, a new, standardisable candle that can reach more distant Universe has been awaited.
Here we propose a new distance measure using fast radio bursts (FRBs), which are a new emerging population of $\sim$ ms time scale radio bursts that can reach high-$z$ in quantity.
We show an empirical positive correlation between the time-integrated luminosity (L$_{\nu}$) and rest-frame intrinsic duration ($w_{\rm int,rest}$) of FRBs.
The L$_{\nu}-w_{\rm int,rest}$ correlation is with a weak strength but statistically very significant, i.e., Pearson coefficient is $\sim$ 0.5 with p-value of $\sim$0.038, despite the smallness of the current sample.
This correlation can be used to measure intrinsic luminosity of FRBs from the observed $w_{\rm int,rest}$.
By comparing the luminosity with observed flux, we measure luminosity distances to FRBs, and thereby construct the Hubble diagram.
This FRB cosmology with the L$_{\nu}-w_{\rm int,rest}$ relation has several advantages over SNe Ia, Gamma-Ray Burst (GRB), and well-known FRB dispersion measure (DM)-$z$ cosmology; (i) access to higher redshift Universe beyond the SNe Ia, (ii) high event rate that is $\sim$ 3 order of magnitude more frequent than GRBs, and (iii) it is free from the uncertainty from intergalactic electron density models, i.e., we can remove the largest uncertainty in the well-debated DM-$z$ cosmology of FRB.
Our simulation suggests that the L$_{\nu}-w_{\rm int,rest}$ relation provides us with useful constraints on the time variability of the dark energy when the next generation radio telescopes start to find FRBs in quantity.
\end{abstract}

\begin{keywords}
(cosmology:) cosmological parameters -- (cosmology:) dark energy  -- radio continuum: transients -- (cosmology:) distance scale
\end{keywords}



\section{Introduction}
\label{introduction}
Revealing nature of dark energy is one of the biggest challenge in astronomy and physics. 
Especially the time variability of dark energy remains unknown. 
Type Ia supernovae (SNe Ia) have been used to map the accelerating expansion of the Universe, reaching as far as $z\sim$1.7 \citep{Riess1998,Suzuki2012}.
However, much farther distances can not be measured by SNe Ia.
The event rate of SNe Ia declines at higher redshifts due to the time delay from star formation \citep{Rodney2014}. 
SNe Ia are too faint to be observed at more distant Universe even with the largest telescopes to date. 
To constrain the equation of state of the dark energy parameterised by $w$ (=P/$\rho$, i.e., pressure divided by energy density), especially the time dependence of $w$, it is essentially important to observe high-$z$ Universe to break the degeneracy between cosmological models \citep{Linder2003a,King2014}. 
In regard to this point, a new, standardisable candle that can reach more distant Universe has been awaited.

Several candidates of high-$z$ standard candle have been proposed. 
Gamma-Ray Bursts (GRBs) show an empirical correlation between an isotropic energy and spectral energy peak \citep{Amati2002,Yonetoku2004} which enables us to estimate an intrinsic luminosity from observed spectral peak energy independent from redshift and cosmological assumptions \citep[e.g.,][]{Tsutsui2009}.
Quasars also could be used as a standard candle due to a correlation between the intrinsic luminosity and X-ray to UV flux ratio \citep{Risaliti2015,Lusso2016,Lusso2017}.
CO galaxies were proposed as a possible standard candle, which show a positive correlation between CO (1-0) line width and the intrinsic luminosity \citep[e.g.,][]{Goto2015,Wu2019}.
Although these candidates have a potential to investigate the high-$z$ Universe beyond SNe Ia, the empirical correlations need to be firmed further in order to use them as a powerful tool to constrain the nature of dark energy.

Fast Radio Bursts (FRBs), which are a new emerging population of $\sim$ ms time scale radio busts, have a potential to reveal the nature of dark energy, because FRBs can reach high-$z$ beyond SNe Ia in quantity \citep{Fialkov2017}.
Indeed the observed dispersion measure (DM) of FRBs can be used to constrain the cosmological parameters in combination with independent measurement of redshift \citep[e.g.,][]{Zhou2014}. DM is an indicator of distance to FRB because the more distant FRB should have the larger DM due to the larger amount of intergalactic medium between the FRB and Milky Way.
However this DM-$z$ cosmology suffers from an uncertainty of intergalactic electron density models, which degenerates with cosmological parameters \citep{Jaroszynski2019}.
 
In this paper we show a positive correlation between a time-integrated luminosity (L$_{\nu}$) and rest-frame intrinsic duration ($w_{\rm int,rest}$) of FRBs. 
We propose to use the correlation to measure intrinsic luminosity of FRB from observed $w_{\rm int,rest}$. 
For this purpose, the L$_{\nu}$-$w_{\rm int,rest}$ relation needs to be confirmed with spectroscopic redshift that is independent from cosmology and to be calibrated with other standardisable candle.
By comparing the intrinsic luminosity with observed flux, we measure luminosity distances to FRBs, and thereby, construct the Hubble diagram.
As we discuss below, this FRB cosmology with the L$_{\nu}-w_{\rm int,rest}$ relation is free from the uncertainty from intergalactic electron density models.

The structure of this paper is as follows.
In Section \ref{sample_selection} we describe our sample selection criteria.
We demonstrate an empirical positive correlation between L$_{\nu}$ and $w_{\rm int,rest}$ of FRBs in Section \ref{results}. 
Possible physical models and applications of the L$_{\nu}-w_{\rm int,rest}$ relation are discussed in Section \ref{discussion} followed by a conclusion in Section \ref{conclusion}. 

When needed, we assume the first-year Wilkinson Microwave Anisotropy Probe (WMAP1) cosmology \citep{Spergel2003} as a fiducial model, i.e., $\Lambda$ cold dark matter cosmology with ($\Omega_{m}$,$\Omega_{\Lambda}$,$\Omega_{b}$,$h$)=(0.27, 0.73, 0.044, 0.71), unless otherwise mentioned.
Throughout this paper, we assume that each burst has the same duration in all radio frequencies.

\section{Sample selection}
\label{sample_selection}
\subsection{Source catalogue}
We compiled 68 \lq \lq verified \rq \rq FRBs from the FRBCAT project\footnote[1]{http://frbcat.org/} \citep{Petroff2016}, which were confirmed via publication, or received with a high importance score over the VOEvent Network.
The compiled catalogue includes FRB ID, telescope, galactic latitude ($b$), longitude ($l$), sampling time ($w_{\rm sample}$), central frequency ($\nu_{\rm obs}$), channel bandwidth ($\Delta \nu_{\rm obs}$), presence of scattering broadening, observed dispersion measure (DM$_{\rm obs}$), observed burst duration ($w_{\rm obs}$), and observed fluence (E$_{\nu_{\rm obs}}$) together with errors of these observed parameters.

\subsection{Deriving redshift and DM$_{\rm IGM}$}
The observed dispersion measure contains the contributions from the Milky Way (DM$_{\rm MW}$), intergalactic medium (DM$_{\rm IGM}$), and host galaxy of FRB (DM$_{\rm host}$).
\begin{equation}
\label{eq1}
{\rm DM}_{\rm obs}={\rm DM}_{\rm MW}(b,l)+{\rm DM}_{\rm IGM}(z)+{\rm DM}_{\rm host}(z).
\end{equation}
To estimate DM$_{\rm MW}$ we used the YMW16 electron-density model \citep{Yao2017} of the Milky May.
In the calculation, DM$_{\rm MW}$ is accumulated up to the distance of 20 kpc along the line of sight to FRBs.
The host galaxy's contribution is assumed as DM$_{\rm host} = 50.0/(1+z)$ pc cm$^{-3}$ by following the previous work \citep{Shannon2018}.
Here DM$_{\rm IGM}$ reflects the amount of electrons in between the FRB host and the Milky Way.
Apart from the local inhomogeneity of the intergalactic medium, the amount of the intervening electrons increases with increasing distance to the FRB on average.
Therefore, DM$_{\rm IGM}$ can be an indicator of distance to FRB or redshift by assuming the cosmology and density evolution.

The mean dispersion measure of intergalactic medium \citep[e.g.,][]{Zhou2014} is expressed as
\begin{equation}
\label{eq2}
\begin{split}
&\mathrm{DM}_{\mathrm{IGM}}(z)=\Omega_{\mathrm{b}} \frac{3 H_{0} c}{8 \pi G m_{\mathrm{p}}} \times \\
&\int_{0}^{z} \frac{\left(1+z^{\prime}\right) f_{\mathrm{IGM}}\left(z^{\prime}\right)\left(Y_{\mathrm{H}} X_{\mathrm{e}, \mathrm{H}}\left(z^{\prime}\right)+\frac{1}{2} Y_{\mathrm{p}} X_{\mathrm{e}, \mathrm{He}}\left(z^{\prime}\right)\right)}{\left\{\Omega_{\mathrm{m}}\left(1+z^{\prime}\right)^{3}+\Omega_{\Lambda}\left(1+z^{\prime}\right)^{3\left[1+w\left(z^{\prime}\right)\right]}\right\}^{1 / 2}} d z^{\prime},
\end{split}
\end{equation}
for the flat Universe, where $X_{\mathrm{e}, \mathrm{H}}$ and $X_{\mathrm{e}, \mathrm{He}}$ are the ionisation fractions of the intergalactic hydrogen and helium, respectively.
$Y_{\mathrm{H}}=\frac{3}{4}$ and $Y_{\mathrm{p}}=\frac{1}{4}$ are mass fractions of H and He.
$f_{\mathrm{IGM}}$ is a fraction of baryons in the Universe contained within intergalactic medium.
The equation of state of dark energy is expressed as $w$.
We assumed $X_{\mathrm{e}, \mathrm{H}}=1$ and $X_{\mathrm{e}, \mathrm{He}}=1$.
This assumption is reasonable up to $z \sim 3$ \citep{Zhou2014}, because intergalactic medium is fully ionised.
We use $f_{\mathrm{IGM}}$ = 0.9 at $z>1.5$ and $f_{\mathrm{IGM}}= 0.053z+0.82$ at $z<1.5$ by following the literature \citep{Zhou2014}.

Thus, Eq. (\ref{eq1}) is a function of redshift with the observed DM$_{\rm obs}$ and calculated DM$_{\rm MW}$. 
The solution provides individual FRBs with redshift and DM$_{\rm IGM}$ estimates.

\subsection{Estimate of intrinsic duration}
The observed pulse duration of FRB is broadened by the instrumental and astrophysical sources in general.
To estimate the intrinsic duration of FRB, these broadening components need to be subtracted.
The instrumental pulse broadening includes the sampling time scale and dispersion smearing.
It is obvious that the temporal resolution can not be better than the sampling time scale, which broadens the pulse.
The dispersion smearing is caused by the finite spectral resolution of radio observations.
The arrival time of FRB delays depending on the frequency due to the dispersion.
This delay still exists in the intra-channel bandwidth which broadens the observed pulse.
The larger the dispersion measure, the larger the time delay.
Thus the broadened width by the dispersion smearing is expressed as
\begin{equation}
\label{eq3}
w_{\rm DS}=8.3 \times 10^{-3} \left( \frac{\rm DM_{\rm obs}}{\rm pc~cm^{-3}} \right)\left(\frac{\Delta \nu_{\rm obs}}{\rm MHz} \right) \left( \frac{\nu_{\rm obs}}{\rm GHz}\right)^{-3} {\rm ms},
\end{equation}
where $\Delta \nu_{\rm obs}$ is the channel bandwidth and $\nu_{\rm obs}$ is the observing frequency.

The astrophysical source of pulse broadening is the scattering process of the radio emission possibly in the intergalactic medium or host galaxy, since the scattering in the Milky Way is negligible \citep{Macquart2013}.
In fact some FRBs show clear asymmetric broadening tails which are considered to be the scattering \citep{Thornton2013}.
Although more data are required to conclude which source predominantly contributes to the scattering, it is likely that the strong turbulence in host galaxies cause such an obvious scattering broadening.
We denote the scattering broadening as $w_{\rm IGM/host}$ to express possible contributions from intergalactic medium and host galaxy.
By assuming a Gaussian function for instrumental broadening, they can be subtracted as follows.
\begin{equation}
\label{eq4}
w_{\rm int+IGM/host}^{2}=w_{\rm obs}^{2}-w_{\rm sample}^{2}-w_{\rm DS}^{2},
\end{equation}
where $w_{\rm int+IGM/host}$ is the convolution between the intrinsic duration of FRB ($w_{\rm int}$) and $w_{\rm IGM/host}$, and $w_{\rm sample}$ is the sampling time scale.

For the FRBs without any scattering feature, Eq.(\ref{eq4}) can be simplified to
\begin{equation}
\label{eq5}
w_{\rm int}^{2}=w_{\rm obs}^{2}-w_{\rm sample}^{2}-w_{\rm DS}^{2},
\end{equation}
given that $w_{\rm IGM/host}$ is negligible.
By following Eq.(\ref{eq3}), (\ref{eq4}), and (\ref{eq5}), $w_{\rm int+IGM/host}$ and $w_{\rm int}$ are calculated.
We note that $w_{\rm int}$ is the intrinsic duration in the observed frame.
The rest frame intrinsic duration, $w_{\rm int, rest}$, is calculated as $w_{\rm int, rest}=w_{\rm int}/(1+z)$.

\subsection{Estimate of time-integrated luminosity}
The time-integrated luminosity of FRB at rest frame $\nu_{\rm rest}$ GHz, L$\nu_{\rm rest}$, is calculated by following literature \citep[i.e., time integration of Eq. 2 in][]{Novak2017}.
\begin{equation}
\label{eq6}
{\rm L}_{\nu_{\rm rest}}=\frac{4\pi d_{l}(z)^{2}}{(1+z)^{2+\alpha}} \left(\frac{\nu_{\rm rest}}{\nu_{\rm obs}} \right)^{\alpha} {\rm E}_{\nu_{\rm obs}},
\end{equation}
where d$_{l}(z)$ is the luminosity distance to FRB calculated from redshift and E$_{\nu_{\rm obs}}$ is the observed fluence.
The WMAP1 cosmology \citep{Spergel2003} is assumed to derive d$_{l}(z)$.
We assumed power law spectra expressed as $E_{\nu_{\rm obs}} \propto \nu_{\rm obs}^{\alpha}$.
Among 27 robust sample (see Section \ref{robust_sample}), $\alpha$ is measured for 15 FRBs \citep{Macquart2019}.
We used the individual $\alpha$ measurement for the 15 FRBs, and assumed a mean value of $\alpha=-1.5$ \citep{Macquart2019} for other FRBs.
The rest frame frequency, $\nu_{\rm rest}=1.83$ GHz, is adopted to minimise the factor of $\frac{1}{(1+z)^{2+\alpha}}\left( \frac{\nu_{\rm rest}}{\nu_{\rm obs}} \right)^{\alpha}$ in Eq. (\ref{eq6}).

\subsection{Selection of robust sample}
\label{robust_sample}
The correction terms of DM$_{\rm MW}$, DM$_{\rm host}$,$w_{\rm sample}$, $w_{\rm DS}$, and pulse broadening by scattering could have large uncertainties to calculate DM$_{\rm IGM}$, i.e., redshift, and $w_{\rm int}$.
Therefore we applied criteria to select robust sample as follows.
\begin{itemize}
\item[(i)] DM$_{\rm IGM}$ $\geq$ 0.5 DM$_{\rm obs}$
\item[(ii)] $w_{\rm int}$ $\geq$ 0.5 $w_{\rm obs}$
\item[(iii)] No scattering feature
\end{itemize}
The criterion (i) ensures the robust estimate of DM$_{\rm IGM}$ and redshift by excluding FRBs that have relatively large correction terms of DM$_{\rm MW}$ and DM$_{\rm host}$.
The criterion (ii) is same as (i) but for $w_{\rm int}$ by excluding FRBs with relatively large instrumental broadening.
For the criterion (iii), we excluded FRBs that show a hint of scattering tail in their pulses from the robust sample except for FRB 180110, 180119, 180130, and FRBs detected by CHIME for which pulse duration are reported after deconvolution of exponential pulse broadening function \citep{Shannon2018,CHIMEFRB2019}.
The pulse broadening by scattering is attributed to the host galaxy or intergalactic medium.
In any case, the scattering is a contamination for $w_{\rm int}$ estimate.
Therefore FRBs indicating scattering are excluded from our analysis.
By applying the criteria, we selected totally 27 robust FRBs.
Observed and derived parameters of the robust sample are summarised in table \ref{tab1} and \ref{tab2}, respectively.

\begin{table*}
	\centering
	\caption{
	Observed parameters of 27 robust FRBs selected from FRBCAT catalogue \citep{Petroff2016}.
    }
	\label{tab1}
	\begin{flushleft}
	\begin{tabular}{|l|c|c|c|c|c|c|c|}\hline
(1)& (2)            & (3)       & (4)             & (5)                   & (6)        & (7)              & (8) \\ \hline
ID  & DM & E$_{\nu_{\rm obs}}$ & $w_{\rm obs}$ & $\alpha$ & $\nu_{\rm obs}$ & $\Delta \nu_{\rm obs}$ & $w_{\rm sample}$ \\
 & (pc cm$^{-3}$)  & (Jy ms)  & (ms) &          & (GHz)           &     (MHz)     &  (ms) \\ \hline 
010125 & 790.00$\pm$3.00 & 2.82$^{+2.99}_{-2.27}$ & 9.40$\pm$2.65 & -1.50 & 1.37 & 3.00 & 0.12 \\
010312 & 1187.00$\pm$14.00 & 6.08$^{+0.33}_{-0.33}$ & 24.30$\pm$1.30 & -1.50 & 1.37 & 3.00 & 1.00 \\
110214 & 168.90$\pm$0.50 & 51.30$^{+2902.70}_{-24.30}$ & 1.90$\pm$0.90 & -1.50 & 1.35 & 0.39 & 0.06 \\
110220 & 944.38$\pm$0.05 & 7.28$^{+0.13}_{-0.13}$ & 5.60$\pm$0.10 & -1.50 & 1.35 & 0.39 & 0.06 \\
110626 & 723.00$\pm$0.30 & 0.56$^{+3.98}_{-0.51}$ & 1.40$\pm$0.83 & -1.50 & 1.35 & 0.39 & 0.06 \\
110703 & 1103.60$\pm$0.70 & 2.15$^{+2.73}_{-1.41}$ & 4.30$\pm$2.04 & -1.50 & 1.35 & 0.39 & 0.06 \\
120127 & 553.30$\pm$0.30 & 0.55$^{+1.04}_{-0.30}$ & 1.10$\pm$0.45 & -1.50 & 1.35 & 0.39 & 0.06 \\
150215 & 1105.60$\pm$0.80 & 2.02$^{+1.98}_{-0.73}$ & 2.88$\pm$0.89 & -1.50 & 1.35 & 0.39 & 0.06 \\
160317 & 1165.00$\pm$11.00 & 63.00$^{+21.00}_{-21.00}$ & 21.00$\pm$7.00 & -1.50 & 0.84 & 0.78 & 0.66 \\
160410 & 278.00$\pm$3.00 & 28.00$^{+7.00}_{-7.00}$ & 4.00$\pm$1.00 & -1.50 & 0.84 & 0.78 & 0.66 \\
170416 & 523.20$\pm$0.20 & 97.00$^{+2.00}_{-2.00}$ & 5.00$\pm$0.60 & -7.50 & 1.32 & 1.00 & 1.26 \\
170428 & 991.70$\pm$0.90 & 33.88$^{+2.00}_{-2.00}$ & 4.40$\pm$0.50 & -2.10 & 1.32 & 1.00 & 1.26 \\
170707 & 235.20$\pm$0.60 & 51.80$^{+3.00}_{-3.00}$ & 3.50$\pm$0.50 & 1.60 & 1.30 & 1.00 & 1.26 \\
170906 & 390.30$\pm$0.40 & 74.00$^{+7.00}_{-7.00}$ & 2.50$\pm$0.30 & -6.30 & 1.30 & 1.00 & 1.26 \\
171004 & 304.00$\pm$0.30 & 44.00$^{+2.00}_{-2.00}$ & 2.00$\pm$0.30 & 2.60 & 1.30 & 1.00 & 1.26 \\
171019 & 460.80$\pm$1.10 & 218.70$^{+5.00}_{-5.00}$ & 5.40$\pm$0.30 & -12.00 & 1.30 & 1.00 & 1.26 \\
171116 & 618.50$\pm$0.50 & 62.72$^{+2.00}_{-2.00}$ & 3.20$\pm$0.50 & 1.70 & 1.30 & 1.00 & 1.26 \\
171216 & 203.10$\pm$0.50 & 39.90$^{+2.00}_{-2.00}$ & 1.90$\pm$0.30 & 2.60 & 1.30 & 1.00 & 1.26 \\
180119 & 402.70$\pm$0.70 & 109.89$^{+3.00}_{-3.00}$ & 2.70$\pm$0.50 & -1.10 & 1.30 & 1.00 & 1.26 \\
180128.0 & 441.40$\pm$0.20 & 50.75$^{+2.00}_{-2.00}$ & 2.90$\pm$0.30 & -2.30 & 1.30 & 1.00 & 1.26 \\
180130 & 343.50$\pm$0.40 & 94.71$^{+3.00}_{-3.00}$ & 4.10$\pm$1.00 & 0.49 & 1.30 & 1.00 & 1.26 \\
180131 & 657.70$\pm$0.50 & 99.90$^{+3.00}_{-3.00}$ & 4.50$\pm$0.40 & -2.40 & 1.30 & 1.00 & 1.26 \\
180212 & 167.50$\pm$0.50 & 95.93$^{+8.00}_{-8.00}$ & 1.81$\pm$0.06 & -3.70 & 1.30 & 1.00 & 1.26 \\
180324 & 431.00$\pm$0.40 & 70.95$^{+3.00}_{-3.00}$ & 4.30$\pm$0.50 & 0.85 & 1.30 & 1.00 & 1.26 \\
180525 & 388.10$\pm$0.30 & 299.82$^{+6.00}_{-6.00}$ & 3.80$\pm$0.10 & -1.30 & 1.30 & 1.00 & 1.26 \\
180727.J1311+26 & 642.07$\pm$0.03 & 14.00$^{+10.00}_{-10.00}$ & $\sim$1.4 & -1.50 & 0.60 & 0.024 & 0.98 \\
180812.J0112+80 & 802.57$\pm$0.04 & 18.00$^{+12.00}_{-12.00}$ & $\sim$1.8 & -1.50 & 0.60 & 0.024 & 0.98 \\ \hline
    \end{tabular}\\
    Column (1) FRB ID. (2) Observed dispersion measure. (3) Observed fluence. If the uncertainty, $\delta$E$_{\nu_{\rm obs}}$, is not provided, we calculated it as $\delta$E$_{\nu_{\rm obs}}$=E$_{\nu_{\rm obs}} \times \sqrt{(\delta w_{\rm obs}/w_{\rm obs})^{2}+(\delta f_{\rm obs}/f_{\rm obs})^{2}}$, where $f_{\rm obs}$ and $\delta f_{\rm obs}$ are observed flux density and the uncertainty, respectively. (4) Observed duration. Observed frame intrinsic durations, $w_{\rm int}$, were reported for FRB 180727.J1311+26 and 180812.J0112+80 after subtracting instrumental broadening and scattering component \citep{CHIMEFRB2019} instead of $w_{\rm obs}$. Here $w_{\rm obs}$ of these two FRBs were calculated based on Eq. (\ref{eq5}). (5) Spectral index, i.e., E$_{\nu_{\rm obs}} \propto \nu_{\rm obs}^{\alpha}$, compiled from literature \citep{Macquart2019}. We assumed a mean value of $\alpha=-1.5$ \citep{Macquart2019}, if $\alpha$ is not available. (6) Observed frequency. (7) Channel width. (8) Observational sampling time interval. \\
    \end{flushleft}
\end{table*}

\begin{table*}
	\centering
	\caption{
	Derived parameters of 27 robust FRBs with two cosmological assumptions.}
	\label{tab2}
	\begin{flushleft}
	\begin{tabular}{|l|c|c|c|c|c|c|c|}\hline
(1)& (2)            & (3)       & (4)             & (5)                   & (6)        & (7)              & (8) \\ \hline
   &                &   \multicolumn{3}{c}{WMAP1} & \multicolumn{3}{c}{Planck15}  \\ 
   \cmidrule(lr){3-5}\cmidrule(lr){6-8}
ID & DM$_{\rm MW}$  & Redshift  & logL$_{\nu}$    & log$w_{\rm int,rest}$ &  Redshift  & logL$_{\nu}$     & log$w_{\rm int,rest}$  \\
   &  (pc cm$^{-3}$)  &           & (erg Hz$^{-1}$)& (ms)                  &            & (erg Hz$^{-1}$) &  (ms) \\ \hline 
010125  &  75.88  &  0.782$\pm$0.081  &  31.59$\pm$0.36  &  0.49$\pm$0.19   &  0.759$\pm$0.080  &  31.59$\pm$0.35  &  0.50$\pm$0.19  \\
010312  &  54.88  &  1.231$\pm$0.098  &  32.37$\pm$0.08  &  0.98$\pm$0.04   &  1.203$\pm$0.096  &  32.37$\pm$0.08  &  0.99$\pm$0.04  \\
110214  &  21.06  &  0.132$\pm$0.048  &  31.14$\pm$0.97  &  0.22$\pm$0.22   &  0.126$\pm$0.045  &  31.13$\pm$1.03  &  0.22$\pm$0.22  \\
110220  &  24.12  &  1.004$\pm$0.089  &  32.24$\pm$0.09  &  0.44$\pm$0.02   &  0.978$\pm$0.087  &  32.23$\pm$0.09  &  0.44$\pm$0.02  \\
110626  &  33.57  &  0.756$\pm$0.079  &  30.85$\pm$0.64  &  -0.23$\pm$0.30   &  0.733$\pm$0.079  &  30.84$\pm$0.62  &  -0.23$\pm$0.30  \\
110703  &  23.08  &  1.176$\pm$0.094  &  31.86$\pm$0.35  &  0.27$\pm$0.27   &  1.148$\pm$0.091  &  31.86$\pm$0.35  &  0.28$\pm$0.26  \\
120127  &  20.63  &  0.584$\pm$0.073  &  30.59$\pm$0.40  &  -0.28$\pm$0.25   &  0.564$\pm$0.072  &  30.58$\pm$0.39  &  -0.28$\pm$0.24  \\
150215  &  262.36  &  0.921$\pm$0.088  &  31.60$\pm$0.25  &  0.11$\pm$0.19   &  0.896$\pm$0.085  &  31.59$\pm$0.26  &  0.12$\pm$0.18  \\
160317  &  394.61  &  0.843$\pm$0.086  &  32.70$\pm$0.18  &  0.96$\pm$0.22   &  0.819$\pm$0.083  &  32.69$\pm$0.19  &  0.97$\pm$0.21  \\
160410  &  56.71  &  0.224$\pm$0.060  &  31.06$\pm$0.24  &  0.32$\pm$0.18   &  0.214$\pm$0.050  &  31.06$\pm$0.24  &  0.32$\pm$0.18  \\
170416  &  27.51  &  0.543$\pm$0.072  &  33.03$\pm$0.25  &  0.46$\pm$0.07   &  0.524$\pm$0.067  &  32.99$\pm$0.24  &  0.47$\pm$0.07  \\
170428  &  27.38  &  1.051$\pm$0.093  &  33.04$\pm$0.10  &  0.04$\pm$0.17   &  1.025$\pm$0.089  &  33.03$\pm$0.10  &  0.04$\pm$0.17  \\
170707  &  26.89  &  0.208$\pm$0.050  &  31.75$\pm$0.17  &  0.41$\pm$0.08   &  0.199$\pm$0.048  &  31.75$\pm$0.16  &  0.42$\pm$0.08  \\
170906  &  26.57  &  0.394$\pm$0.066  &  32.28$\pm$0.26  &  0.05$\pm$0.14   &  0.378$\pm$0.061  &  32.25$\pm$0.25  &  0.06$\pm$0.14  \\
171004  &  32.98  &  0.285$\pm$0.056  &  31.93$\pm$0.11  &  -0.09$\pm$0.17   &  0.273$\pm$0.054  &  31.94$\pm$0.11  &  -0.09$\pm$0.17  \\
171019  &  26.28  &  0.474$\pm$0.070  &  33.16$\pm$0.36  &  0.53$\pm$0.04   &  0.457$\pm$0.065  &  33.09$\pm$0.33  &  0.53$\pm$0.04  \\
171116  &  37.48  &  0.637$\pm$0.074  &  32.50$\pm$0.05  &  0.03$\pm$0.18   &  0.616$\pm$0.072  &  32.51$\pm$0.05  &  0.04$\pm$0.16  \\
171216  &  28.67  &  0.166$\pm$0.048  &  31.55$\pm$0.18  &  0.01$\pm$0.17   &  0.158$\pm$0.047  &  31.56$\pm$0.19  &  0.01$\pm$0.16  \\
180119  &  37.88  &  0.395$\pm$0.063  &  32.48$\pm$0.14  &  0.12$\pm$0.15   &  0.379$\pm$0.059  &  32.47$\pm$0.14  &  0.12$\pm$0.16  \\
180128.0  &  26.56  &  0.452$\pm$0.080  &  32.28$\pm$0.16  &  0.14$\pm$0.10   &  0.435$\pm$0.063  &  32.27$\pm$0.15  &  0.14$\pm$0.10  \\
180130  &  26.14  &  0.340$\pm$0.070  &  32.31$\pm$0.13  &  0.44$\pm$0.14   &  0.326$\pm$0.057  &  32.31$\pm$0.13  &  0.44$\pm$0.15  \\
180131  &  26.90  &  0.692$\pm$0.079  &  33.06$\pm$0.13  &  0.32$\pm$0.07   &  0.670$\pm$0.074  &  33.04$\pm$0.12  &  0.32$\pm$0.07  \\
180212  &  27.83  &  0.121$\pm$0.045  &  31.09$\pm$0.35  &  0.00$\pm$0.04   &  0.116$\pm$0.040  &  31.08$\pm$0.34  &  0.01$\pm$0.04  \\
180324  &  71.16  &  0.389$\pm$0.061  &  32.29$\pm$0.11  &  0.43$\pm$0.07   &  0.374$\pm$0.061  &  32.29$\pm$0.11  &  0.44$\pm$0.07  \\
180525  &  27.48  &  0.390$\pm$0.062  &  32.90$\pm$0.15  &  0.37$\pm$0.02   &  0.375$\pm$0.059  &  32.90$\pm$0.15  &  0.38$\pm$0.02  \\
180727.J1311+26  &  19.78  &  0.683$\pm$0.078  &  31.62$\pm$0.30  &  -0.33$\pm$0.09   &  0.661$\pm$0.077  &  31.61$\pm$0.31  &  -0.33$\pm$0.09  \\
180812.J0112+80  &  93.39  &  0.777$\pm$0.081  &  31.85$\pm$0.30  &  -0.15$\pm$0.18   &  0.754$\pm$0.078  &  31.85$\pm$0.31  &  -0.15$\pm$0.18  \\ \hline
    \end{tabular}\\
    Column (1) FRB ID. (2) Dispersion measure of Milky Way along a line of sight to FRB, based on YMW16 model \citep{Yao2017}. DM$_{\rm MW}$ is accumulated up to 20 kpc. (3) and (6) Redshift calculated from Eq. (\ref{eq1}). (4) and (7) Time-integrated luminosity at rest frame 1.83 GHz. (5) and (8) Rest frame intrinsic duration. All uncertainties include observational uncertainties of pulse duration, dispersion measure, and fluence together with a density fluctuation of intergalactic medium estimated from a simulation \citep{Zhu2018} (see Section \ref{sample_selection} for details).
    \end{flushleft}
\end{table*}

\section{Results}
\label{results}
\subsection{Luminosity-Duration Relation}
\begin{figure*}
	\includegraphics[width=\columnwidth]{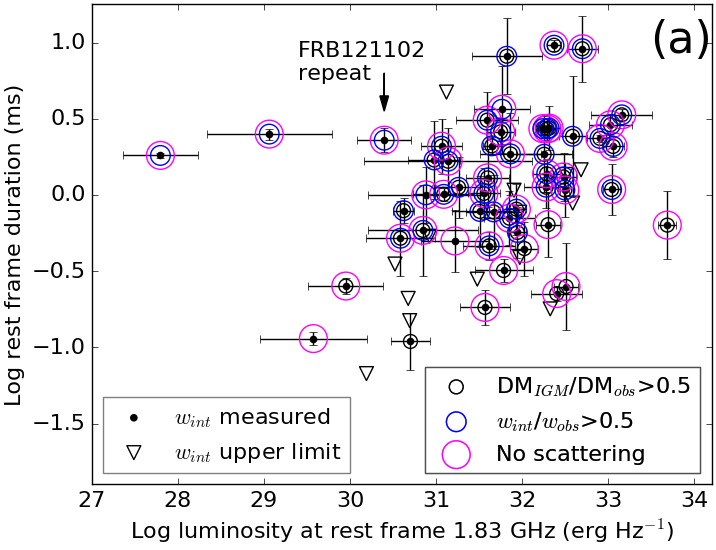}
	\includegraphics[width=\columnwidth]{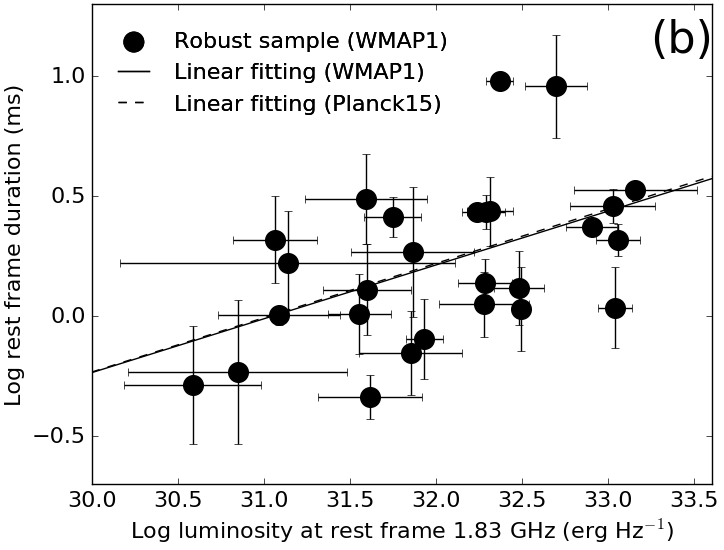}
    \caption{The luminosity-duration relation of FRBs at rest frame 1.83 GHz.
    (left) 
    FRBs with the measured $w_{\rm int}$ are displayed with black dots, while upper limits are with open triangles.
    We applied three criteria to all the FRBs in the left panel to construct a robust sample in the right panel.
    FRBs with each criteria are marked by different coloured circles.
    FRBs with the robust DM$_{\rm IGM}$ measurements are marked with black circles.
    Blue circles show FRBs with the robust $w_{\rm int}$ measurements.
    Magenta circles are FRBs without any feature of the scattering broadening in their pulses.
    (right)
    The robust sample that satisfies all three criteria is shown in the right panel with black dots.
    These data assume WMAP1 cosmology.
    The best fit of the linear function is shown with the black solid line.
    When the assumed cosmology is changed, the data points move slightly. However, for clarity, we only show the best fit line for another cosmology of Planck15 \citep{Planck15}, i.e., ($\Omega_{m}$,$\Omega_{\Lambda}$,$\Omega_{b}$,$h$)=(0.307, 0.693, 0.0486, 0.677).
    Error bars include observational uncertainties of pulse duration, dispersion measure, and fluence together with a density fluctuation of intergalactic medium estimated from a simulation \citep{Zhu2018} (see Section \ref{sample_selection} for details).
    }
    \label{fig1}
\end{figure*}

As described in Section \ref{sample_selection}, we selected 27 robust FRBs from 68 confirmed FRBs in the FRBCAT project \citep{Petroff2016}.
The robust FRBs must have secure measurements of the dispersion measure of intergalactic medium (DM$_{\rm IGM}$) and intrinsic duration ($w_{\rm int}$) without any sign of the scattering broadening (Fig. \ref{fig1}a).
The scattering broadening is considered to be caused by the turbulence in the host galaxy or intergalactic medium.
Since in either case this effect is a contamination in measuring the intrinsic duration of FRBs. 
Based on the FRBCAT catalogue, FRBs with a sign of the scattering broadening were excluded from the robust sample (see Section \ref{sample_selection} for details).

At the moment, most FRBs do not have observed redshifts.
Therefore, to measure the L$_{\nu}$, the robust measurement of DM$_{\rm IGM}$ was used to estimate redshifts of FRBs based on Eq. (\ref{eq1}).
L$_{\nu}$ and $w_{\rm int, rest}$ were calculated from the observed fluence, spectral index, pulse duration, and redshift (see Section \ref{sample_selection} for details) with an assumed cosmology.
To examine the L$_{\nu}-w_{\rm int,rest}$ relation, the cosmological parameters of WMAP1 \citep{Spergel2003} are assumed.

In Fig. \ref{fig1}b, the robust FRB sample shows a correlation between L$_{\nu}$ and $w_{\rm int,rest}$.
The best-fit linear function is
\begin{equation}
\label{eq7}
\log (w_{\rm int,rest})=(0.22\pm0.06)\times (\log ({\rm L}_{\nu}) -32.5)+(0.33\pm0.03)
\end{equation}
with a 0.28 dex dispersion from the best fit along $\log (w_{\rm int,rest})$ axis.
While the different cosmological assumptions can affect the L$_{\nu}-w_{\rm int,rest}$ relation, the uncertainty changes the relation only in a systematic way.
The L$_{\nu}-w_{\rm int,rest}$ relation slightly changes when different cosmological parameters are assumed (Fig. \ref{fig1}b).
Regardless of the different cosmological assumptions, the L$_{\nu}-w_{\rm int,rest}$ relation is persistent.

\subsection{Error budget}
To estimate errors reflected in the L$_{\nu}$-$w_{\rm int,rest}$ relation, we included observational uncertainties of pulse duration, dispersion measure, and fluence from the FRBCAT project \citep{Petroff2016}.
The remaining uncertainty is redshift error. 
Since redshift is estimated from Eq. (\ref{eq1}) through DM$_{\rm IGM}$ (Eq. \ref{eq2}), 
the density fluctuation of intergalactic medium is responsible for the redshift uncertainty.
We used the recent simulation results of the dispersion measure of intergalactic medium \citep{Zhu2018}.
In the literature, the stochastic fluctuation of dispersion measures along different lines of sight depends on not only redshift but also the resolution of calculation.
As a conservative assumption, we used the highest uncertainty of dispersion measure as a function of redshift reported in the literature.
Since the error propagation is complicated, we performed Monte Carlo simulations to estimate errors in the L$_{\nu}$-$w_{\rm int,rest}$ relation by independently assigning 10,000 errors to DM$_{\rm obs}$, $w_{\rm obs}$, DM$_{\rm IGM}(z)$, and E$_{\nu_{\rm obs}}$.
In the Monte Calro simulations, the variation of cosmological parameters are not included, since the uncertainty results in the systematic differences in the L$_{\nu}$-$w_{\rm int,rest}$ relation rather than the statistical uncertainty.
Instead, we checked the systematics in Fig. \ref{fig1}b with the different cosmological assumptions.

\begin{figure}
	\includegraphics[width=\columnwidth]{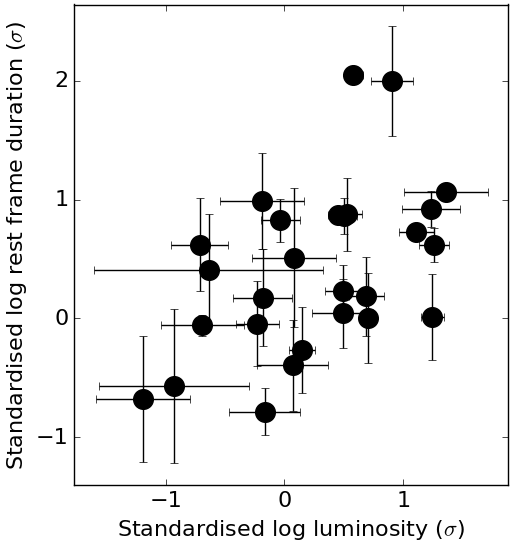}
    \caption{
    The standardised luminosity-duration relation of FRBs.
    Same as Fig. \ref{fig1}b except for standardised axes.
    }
    \label{fig2}
\end{figure}

To examine the dominant source of the dispersion around the L$_{\nu}$-$w_{\rm int,rest}$ relation, Fig. \ref{fig1}b was standardised into Fig. \ref{fig2}.
In Fig. \ref{fig2} the physical scale lengths in both axes are identical in unit of standard deviations of $\log(w_{\rm int, rest})$ and $\log({\rm L}_{\nu})$.
In the standardised space the error of $w_{\rm int, rest}$ is larger than that of L$_{\nu}$ in most cases, which suggests the data dispersion around the L$_{\nu}$-$w_{\rm int,rest}$ relation is dominated by the observational uncertainty of $w_{\rm int,rest}$.
Therefore when the correlation is fitted by a linear function, we use $\log(w_{\rm int,rest})$ as a vertical axis and minimised the residual from the fit function along the $\log(w_{\rm int,rest})$ axis.
The data dispersion around the best-fit linear relation along $\log(w_{\rm int,rest})$ axis is 0.28 dex in physical scale.
This value is actually comparable to a typical error size of $\log(w_{\rm int,rest})$ ($\sim$ 0.2 dex) in Fig. \ref{fig1}b.

\subsection{Pearson coefficient and Kendall's tau}
\begin{figure*}
	\includegraphics[width=\columnwidth]{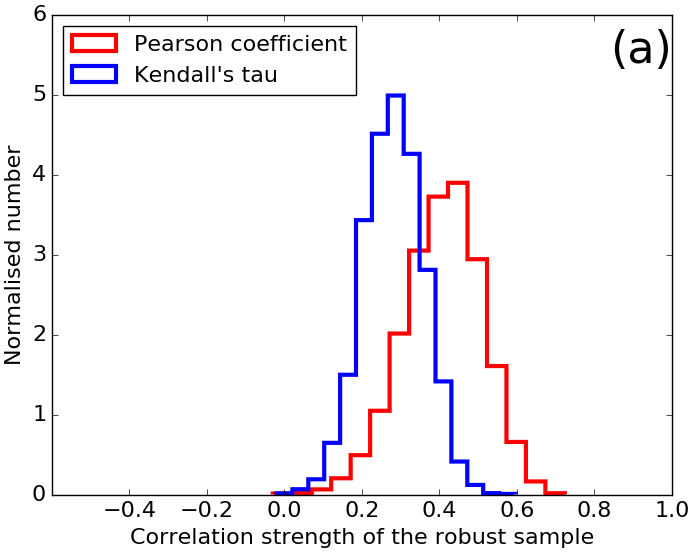}
	\includegraphics[width=\columnwidth]{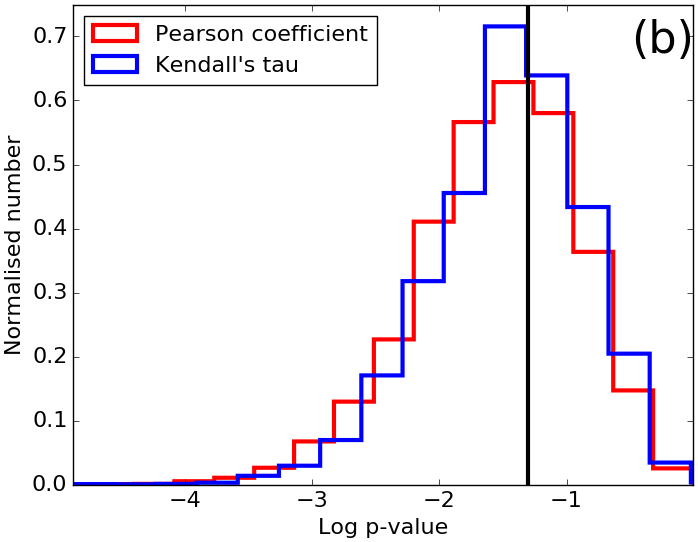}
    \caption{
    The distribution of correlation strength and p-value of the robust sample.
    (left) 
    In the Monte Calro simulation, individual data points in Fig. \ref{fig1}b have 10,000 random errors that follow the Gaussian probability distribution function. 
    The errors of individual data are used for $\sigma$ of the probability distribution functions.
    The correlation strengths calculated from Pearson coefficient and Kendall's tau are shown by red and blue histograms, respectively.
    Note that a Pearson coefficient close to 1 indicates a strong positive correlation while -1 indicates a strong negative correlation.
    A value of Kendall's tau close to 1 indicates strong agreement with a correlation, while -1 is strong disagreement.
    (right) 
    The distribution of p-value. 
    The peaks of the Pearson and Kendall's tau p-values are $p=0.038$ and 0.023 ,respectively.
    A vertical line corresponds to a threshold value of $p=0.05$.
    }
    \label{fig3}
\end{figure*}

The strength of the correlation between the luminosity and intrinsic duration of FRBs was investigated by the Pearson coefficient.
Monte Carlo simulations were performed to take into account errors of individual data in Fig.\ref{fig1}b by adding random errors that follow Gaussian probability distribution function.
We simulated 10,000 errors for individual data and calculated Pearson coefficients (Fig.\ref{fig3}a).
The coefficient distribution peaks around 0.5 that indicates a weak strength of the positive correlation.
Although within the uncertainties of data points the coefficient can be 0.0 that corresponds to no correlation between two parameters, such probability is extremely low.
Supposing that the possible selection bias (see Section \ref{discussion}) weakens the correlation, the coefficient might be the lower limit.
P-value of the Pearson coefficient is an indicator of a statistical significance of correlation.
The p-value peaks at $p=0.038$ (Fig. \ref{fig3}b).
This value is lower than a threshold value of $p=0.05$.
The correlation strength was also checked with Kendall's tau value \citep{Knight1966}.
The value close to 1 indicates strong agreement with a correlation while -1 indicates strong disagreement.
The peak of tau is at $\sim$0.3 indicating a weak strength.
The p-value peaks at 0.023 that is consistent with the Pearson p-value.
Therefore we conclude that the correlation is with a weak strength but statistically significant within observational uncertainties.

\section{Discussion}
\label{discussion}
\subsection{Selection bias}
We imposed three criteria to construct a robust sample as described in Section \ref{sample_selection}.
Here we mention possible biases caused by the criteria.
The criterion (i) preferentially selects FRBs in a large altitude from the Galactic plane.
This is unlikely a bias, because these FRBs are supposed to be extra-galactic sources.
The possible bias of the criterion (i) is that the lower-redshift FRBs have a higher probability to be excluded due to the relatively larger contribution of DM$_{\rm MW}$ and/or DM$_{\rm host}$ to DM$_{\rm obs}$, since DM$_{\rm IGM}$ term is relatively small compared with other terms. 
The low-$z$ FRBs also could be missed due to a small survey volume.
Therefore, in terms of luminosity bias, fainter FRBs might be missed in Fig. \ref{fig1}b even though more fainter FRBs can be detected at lower redshift.
The criterion (ii) tends to exclude intrinsically shorter duration or lower redshift due to a factor of $(1+z)$ broadening.
These biases caused by criteria (i) and (ii) do not significantly affect on the correlation between the luminosity and duration, because only the faint and short end of the relation is limited by the possible biases.
The criteria (iii) could miss a possible contribution from the scattering broadening for faint FRBs, because faint FRBs might not show any clear scattering features because of the lower signal-to-noise ratio even if they actually have scattering broadening.
In this sense fainter FRBs potentially have a scattering contamination that broaden the calculated intrinsic duration.
This possible bias tends to weaken the correlation.

Although there could be selection biases, the impact is unlikely significant or even weakens the correlation.
Therefore it is possible that the less biased future observations might reveal a stronger correlation between the luminosity and duration.

\subsection{Theoretical models to naturally explain the L$_{\nu}$-w$_{\rm int,rest}$ relation}
Astrophysically, the relation will help us understand the physical origin of FRB. 
There exist plethora of physical models of FRBs, but at least one of the models should be able to explain the L$_{\nu}-w_{\rm int,rest}$ relation. 
Models whose luminosity scales with duration would be preferred. 
The models which do not reproduce the relation cannot be the explanation for the all of the FRBs. 
To help theoretical understanding, it is an important task on the observational side to quantify the slope and scatter of the relation more precisely, providing a qualitative test for theoretical models.

A number of physical origins or progenitors of FRBs have been proposed to explain energetic and emission mechanisms for repeating and non-repeating FRBs.
However, there has been no single generally accepted model.
Although it is out of the scope of this paper to fully address the origin of FRBs, we briefly mention individual models for non-repeating FRBs and possible explanations of the L${\nu}-w_{\rm int,rest}$ relation, since repeating FRB 121102 does not meet our selection criteria for the robust sample.
Non-repeating FRB models can be categorised into roughly 5 types; AGN, collapse, collision, merger, and SN remnant.

The AGN model considers an interaction of AGN jet with a surrounding cloud resulting in strongly beamed Bremsstrahlung radiation in pulses \citep{Romero2016}.
The collapse model includes a collapse of a neutron star to a black hole \citep{Falcke2014,Fuller2015}.
During the collapse, a violent magnetic reconnection is expected to release an energy that is large enough to explain FRBs.
The collision model has a variety of FRB progenitors.
One model is a collision between a neutron star and a comet or an asteroid \citep{Geng2015}.
When the neutron star collides with the comet that is confined to the neutron star's poles by strong magnetic stresses, a gravitational energy will be released and an electrostatic equilibrium will be disturbed that results in magnetic reconnection at the collision site to generate an FRB.
The merger model also has a diversity including mergers between any two of a neutron star, a black hole, and a white dwarf.
In the case of the double neutron stars, magnetic breaking is expected to generate coherent radiation in which the standard strength of magnetic field can explain the observed FRB fluxes \citep{Totani2013}.
The SN remnant model considers a neutron star or magnetar surrounded by a nebula.
For the magnetar case, a magnetar flare could reach the nebula.
The interaction between the flare and nebula forms a magnetised shock which generates a synchrotron maser emission, i.e., FRB \citep{Lyubarsky2014}.

Due to the following arguments, the L$_{\nu}-w_{\rm int,rest}$ relation might favour AGN, comet collision, and SN remnant with a magnetar cases.
In the AGN model, the FRB duration is determined by the size of surrounding cloud \citep{Romero2016}.
The volume of the cloud determines the FRB power, resulting in a  positive correlation between the duration and time-integrated luminosity.
The comet collision model predicts that the duration is relevant to the comet mass, which also determines the released gravitational energy \citep{Geng2015}.
The SN remnant with a magnetar model expects that the duration of pulse flare from the magnetar determines the maser pulse duration, i.e., FRB duration.
The total energy of the maser radio emission increases with increasing the pulse flare duration \citep{Lyubarsky2014}.

The slope of the L$_{\nu}-w_{\rm int,rest}$ relation also provides an important hint to address physical models, since different models predict different slopes.
According to the literature \citep{Romero2016,Geng2015,Lyubarsky2014}, the SN remnant with a magnetar model predicts a slope of 1.0.
Slopes of the AGN and comet collision models are $\sim$ 0.3 and 0.4, respectively.
The observed slope of 0.22$\pm$0.06 is closer to the AGN and comet collision models than that of the SN remnant model.
More precise future measurement of the slope will provide us with more stringent constraints on the physical models of FRBs.

Since the current observational data are too small to conclude the physical origin of FRBs, the existence of the L$_{\nu}-w_{\rm int,rest}$ relation does not strictly rule out any other models.
Even the alien civilisations' theory of FRBs cannot be rejected by this argument, because the correlation between L$_{\nu}$ and $w_{\rm int,rest}$ is expected with a slope of $\sim$ 0.3 according to the literature \citep{Lingam2017}.
We need to wait for the improved data to conclude.

\subsection{Possible applications to the cosmology}
Advance in physical understanding behind SN Ia and well explored empirical relations enabled us to use SN Ia as an useful standardisable candle. 
To utilise the L$_{\nu}$-$w_{\rm int,rest}$ relation in cosmological tests as an important ingredient, the relation needs to be well established in both aspects of observation and theory. 
In this sense, not only high quality observations but also revealing the origin of FRBs is necessary to support the physical motivation as a standardisable candle.
If confirmed with future data, the newly discovered L$_{\nu}-w_{\rm int,rest}$ relation will open new possibilities on FRB cosmology. In addition to $z$, and DM, the FRBs will have the 3rd parameter to estimate distances.

(i) A new distance estimate: Using the relation, one can estimate luminosity distance from the observed $w_{\rm int,rest}$. 
By comparing the luminosity distance with the measured redshift (from an optical counterpart, or radio emission lines), one can create a Hubble diagram, and thereby estimate cosmological parameters. 
As estimated in the literature \citep{Zhou2014,Fialkov2017}, near future radio telescopes can easily detect high-$z$ FRB ($z<3$) in quantity. 
And thus, FRBs can be another useful cosmological tool.
It has been pointed out that FRBs can be used for cosmology using DM-$z$ relation \citep{Zhou2014}. 
However, it has been often discussed that it is also difficult to achieve the precision due to the large uncertainties from the scatter and evolution in electron density \citep{Jaroszynski2019}.
In contrast, an important aspect of our method is that it does not require DM-$z$ model, removing one of the largest uncertainties from the FRB cosmology.

(ii) The relation can be also used to probe DM($z$) itself. 
The luminosity distance can be derived from the  L$_{\nu}-w_{\rm int,rest}$ relation. Then, by comparing luminosity distance with the observed DM, then one can map out the DM as a function of the luminosity distance, thereby constraining models on DM($z$) \citep[e.g.,][]{Ioka2003}. 
Observationally, it is more difficult to measure redshifts of FRBs. 
FRBs disappear quickly. 
Locating FRB position is not straightforward. 
However, even when observational redshifts are missing, DM can be mapped out as a function of luminosity distance using our method.


\subsection{Constraints on cosmology}
\label{cosmology}
Regarding (i), we demonstrate how the L$_{\nu}-w_{\rm int,rest}$ relation of FRBs will constrain the cosmological parameters in the future.
In 2020s, the Square Kilometre Array\footnote[2]{https://www.skatelescope.org} (SKA) is expected to be in the scientific operation.
The SKA will be one of the most powerful instruments to detect FRBs thanks to the high sensitivity, the precise localisation, and the wide field of view.
The SKA will be able to discover $\sim 10^{5}$ sky$^{-1}$ yr $^{-1}$ FRBs \citep{Fialkov2017} with secure localisation.
In order to measure spectroscopic redshifts of FRBs, a strategy of follow-up observations to detect host galaxy is needed as carried out for GRB.
In the case of GRBs, host galaxies of $\sim$ 14\% of localised GRBs have been identified with spectroscopic redshifts \footnote[3]{http://www.mpe.mpg.de/~jcg/grbgen.html}\footnote[4]{http://www.grbhosts.org/}.
Therefore, it would not be too optimistic to expect $\sim 10^{3}$ FRB sample with spectroscopic redshifts in the SKA era through optical follow-up observations \citep{Zhou2014}.
Such a sample is large enough to calibrate the L$_{\nu}-w_{\rm int,rest}$ relation to a similar accuracy to the other standard candles such as SNe Ia.
We take the calibrated L$_{\nu}-w_{\rm int,rest}$ relation with spectroscopic redshifts as an input for the simulation.

First, we created $10^{3}$ artificial FRBs with true redshifts between 0 and 3 by following the literature \citep{Zhou2014} and true rest-frame intrinsic durations that follows the distribution of $w_{\rm int,rest}$ found in our robust sample.
Second, the time-integrated luminosity was calculated from the rest-frame intrinsic duration by using Eq. (\ref{eq1}), given that the equation is calibrated.
The possible intrinsic variation of FRBs might contribute to the dispersion of the L$_{\nu}-w_{\rm int,rest}$ relation and then variation of luminosity.
This effect was taken into account when we simulated the observation later.
In the simulation, we assumed $w$CDM with WMAP1 parameters as a true cosmology to calculate the fluence of FRBs from the redshift and time-integrated luminosity.
Here the mean value of observed FRB spectral index, $\alpha=-1.5$ \citep{Macquart2019}, is assumed.
Third, observational uncertainties were assigned to fluence, observed-frame duration and redshift of individual FRBs.
We used the expected sensitivity and dispersion smearing of the SKA (1 mJy and 0.01 ms) for fluence and duration errors \citep{Torchinsky2016}.
The uncertainty of spectroscopic redshifts, $0.00054(1+z)$, is referred from the literature \citep{Scodeggio2018}.

In the future, due to SKA's small observational errors, which are roughly two orders of magnitude smaller than those of current observations, and spectroscopic redshift measurements that free us from the uncertainty of the electron density models of intergalactic medium, the observational dispersion of the L$_{\nu}-w_{\rm int,rest}$ relation will be significantly reduced from the current measurement of 0.28 dex.
In such an era, the dispersion of the L$_{\nu}-w_{\rm int,rest}$ relation may be dominated by the intrinsic variation of FRBs rather than the observational uncertainty.
To reflect the possible intrinsic variation, we assumed two different intrinsic dispersions, i.e., 5\% and 1\% of the dispersion estimated for our robust sample.

Here, the 1\% dispersion case corresponds to a dispersion determined by the observational uncertainty of the SKA, while 5\% case is comparable to a level of the intrinsic luminosity dispersion of SNe Ia as described below.

Having these simulated observational data, we return to the calibrated L$_{\nu}-w_{\rm int,rest}$ relation and estimate time-integrated luminosity from the observed duration of FRBs.
The time-integrated luminosity and observed fluence enable us to re-construct the luminosity distance and distance modulus independently from redshifts.
We utilised the distance modulus as a function of redshift to constrain the cosmological parameters by minimising
\begin{equation}
\chi^{2}={\sum_{i=1}^{n}} \left[ \frac{\mu_{i}^{obs}(z_{i})-\mu_{i}^{th}(z_{i})}{\sigma_{\mu_{i}^{obs}}} \right]^{2},
\end{equation}
where $\mu_{i}^{obs}$ and $\mu_{i}^{th}$ are observed and theoretical distance module of $i$th data.
$\sigma_{\mu_{i}^{obs}}$ is the observational error of $\mu_{i}^{obs}(z_{i})$.
Since we focus on how FRBs reduce the cosmological uncertainties, we compared the following three data sets (i) SNe Ia Union 2.1 \citep{Suzuki2012} only, (ii) SNe Ia and FRBs with 5\% dispersion of the L$_{\nu}-w_{\rm int,rest}$ relation, and (iii) SNe Ia and FRBs with 1\% dispersion with two different cosmological models of $w$CDM and CPL \citep{Chevallier2001,Linder2003b} with a flat universe assumption.
The $w$CDM model assumes a cosmological constant parameterised by $w_{0}$, while the CPL model assumes redshift dependence of dark energy described as $w=w_{0}+w_{a}z/(1+z)$, where $w$ determines the equation of state of the dark energy, i.e., $w=P/\rho$.

\begin{figure*}
	\includegraphics[width=\columnwidth]{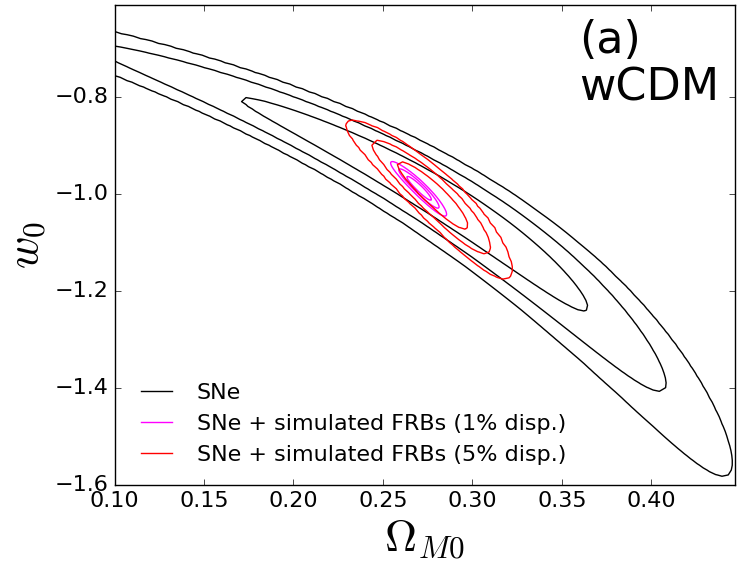}
	\includegraphics[width=\columnwidth]{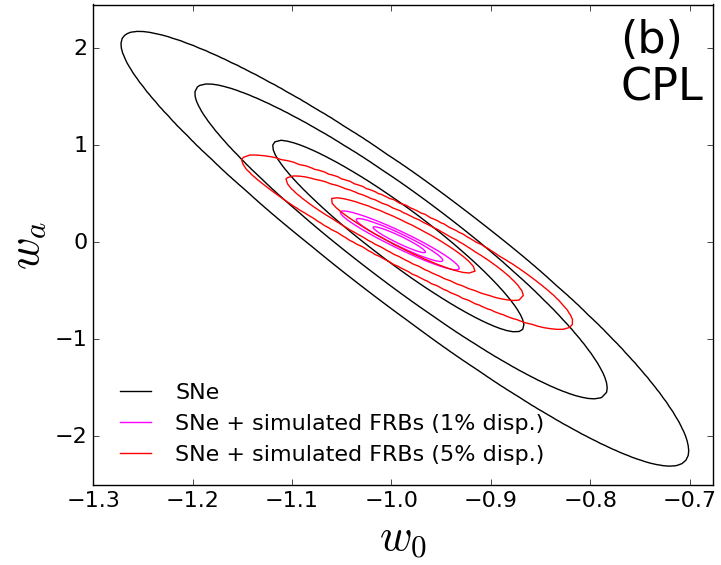}
    \caption{
    The expected constraints on cosmological parameters by the luminosity-duration relation of FRBs.
    (left) 
    Cosmological parameters of $\Omega_{\rm M0}$ and $w_{0}$ of $w$CDM model constrained by SNe Ia (black), SNe Ia/simulated FRBs with 5\% dispersion of the L$_{\nu}-w_{\rm int,rest}$ relation (red), and SNe Ia/simulated FRBs with 1\% dispersion (magenta). Three contours in each colour correspond to 68.27, 95.45, and 99.73\% confidence levels.
    (right) 
    The same as left except for CPL model.
    }
    \label{fig4}
\end{figure*}

The constrained cosmological parameters are shown in Fig. \ref{fig4}.
The different colours correspond to the different data sets and three contours of the same colour are 68.27, 95.45, and 99.73\% confidence levels, where a likelihood function of $\mathscr{L} \propto \exp(-\chi^{2}/2)$ is adopted.
We found that the constraint is essentially limited by the intrinsic dispersion of the L$_{\nu}-w_{\rm int,rest}$ relation.

First, we consider the intrinsic dispersion of 5$\%$ of the currently observed dispersion. This case roughly corresponds to the level of intrinsic luminosity dispersion measured for SNe Ia.
The dispersion of intrinsic luminosity of SNe Ia is $\sim$0.15 mag \citep{Pan2014,Mohlabeng2014} that is 0.06 dex in log luminosity.
If this value is applied to FRBs, 0.06 dex dispersion of intrinsic luminosity is 0.013 dex dispersion of duration, which is $\sim$5\% of the currently observed one.
By combining with SNe Ia, FRBs still provide strong constraints thanks to the higher detection rate and higher maximum redshifts.

Next, we consider the intrinsic dispersion of 1$\%$ of the currently observed dispersion. 
To investigate the 1$\%$ intrinsic dispersion case, the observational uncertainty also needs to be reduced down to the same level.
The 1$\%$ of the current observational uncertainty of $\log w_{\rm int,rest}$ is $0.28 \times 0.01=0.0028$ dex, which corresponds to $\sim$ 0.6\% accuracy.
The expected temporal resolution of the SKA, $\sim 0.01$ ms, is 0.5$\%$ of the typical intrinsic duration of the robust sample, 2 ms.
Therefore observational uncertainty is small enough to investigate the intrinsic dispersion down to the 1\% dispersion case in the SKA era.

If the intrinsic dispersion is 1$\%$ of the currently observed dispersion, FRBs would provide a strong constraint on the dark energy in combination with SNe Ia.
We can constrain the equation of state to $w_{a}\pm 0.13$ and $w_{0}\pm 0.03$ for CPL model, which correspond 26\% and 3\% uncertainties if $w_{a}=0.5$ and $w_{0}=-1.0$, respectively.
These constraints are probably sufficient to figure out the existence of the redshift evolution of the dark energy with an extremely small uncertainty.
In either case, the discovered L$_{\nu}-w_{\rm int,rest}$ relation will open new possibilities in the future FRB sciences.

\subsection{Intrinsic dispersion of the L$_{\nu}$-$w_{\rm int,rest}$ relation}
\label{Intrinsic_dispersion}
We assumed 1 and 5\% of the observed dispersion of the L$_{\nu}$-$w_{\rm int,rest}$ relation in Section \ref{cosmology}. 
The assumptions could be optimistic if the FRB luminosity is controlled by not only duration but also other unknown factors. 
Two decades ago, the peak luminosity of SN Ia has been reported with a dispersion of $\pm$0.8 mag in B band \citep{Phillips1993}. 
So far substantial efforts were made to reduce the dispersion by introducing new correction terms such as light curve shape, colour, and host properties of SN Ia. 
Such detailed corrections allowed us to utilise SN Ia as a standardisable candle. 
To achieve 5\% dispersion case of FRBs, i.e., SN Ia level of the intrinsic luminosity dispersion, it might be necessary to explore additional parameters that determine FRB luminosity. 
A huge number of FRB detections are expected by the next generation telescopes such as CHIME and SKA, e.g., $\sim$10$^{5}$ sky$^{-1}$ yr$^{-1}$ \citep{Fialkov2017} by SKA. 
In terms of statistical number, FRBs would have an advantage in parameter search to reduce the intrinsic luminosity dispersion compared with other possible candidates of high-z standardisable candles such as GRBs \citep[e.g.,][]{Tsutsui2009}.

\subsection{Advantage of the L$_{\nu}$-$w_{\rm int,rest}$ cosmology}
\label{advantage}
As shown in Eq. (\ref{eq2}), the amount of ionised material between FRB and us determines DM($z$).  
The amount of ionised material depends on (i) cosmological parameters, (ii) re-ionisation history of IGM, and (iii) how much material in the Universe are in intergalactic medium phase, i.e., $f_{\rm IGM}$. 
Therefore redshift derived from DM, $z_{\rm DM}$, needs a cosmological assumption. 
The $z_{\rm DM}$ can not be used when cosmological parameters are constrained, because $z_{\rm DM}$ is derived with a cosmological assumption. 
In this sense both of the L$_{\nu}$-$w_{\rm int,rest}$ and DM-$z$ cosmologies need \lq direct\rq\ redshift measurements, i.e., $z_{\rm spec}$.

The difference between the two is whether or not the method uses DM when constraining cosmological parameters. 
DM-$z$ cosmology uses DM and $z_{\rm spec}$. 
The DM includes (ii) and (iii). 
The factors of (ii) and (iii) degenerate with (i), when the DM-$z$ method constrains cosmology \citep[e.g.,][]{Jaroszynski2019}. 
In contrast, the L$_{\nu}$-$w_{\rm int,rest}$ cosmology uses the luminosity derived from duration and $z_{\rm spec}$ to construct the Hubble diagram. 
These parameters are free from above (i), (ii), and (iii), once the L$_{\nu}$-$w_{\rm int,rest}$ relation is confirmed with $z_{\rm spec}$ and is calibrated with other standardisable candle as SN Ia was. 

\section{Conclusions}
\label{conclusion}
We found the empirical positive correlation between the time-integrated luminosity (L$_{\nu}$) and the rest-frame intrinsic duration ($w_{\rm int,rest}$) of FRBs. In the forthcoming future, FRBs are expected to be found in quantity at high-$z$. 
In such an era, this correlation may be useful for cosmology, adding an additional value to FRBs as a new standard candle. 
Notably, the method is free from the uncertainty on the intergalactic electron density evolution that previous FRB's DM-$z$ cosmology suffered from.
Our simulation suggests that the L$_{\nu}-w_{\rm int,rest}$ relation provides us with useful constraints on the time variability of the dark energy when the next generation radio telescopes start to find FRBs in quantity.
The relation will potentially open a new pathway to shed light on the dark energy. 

\section*{Acknowledgements}
We thank R. M. Shannon for kindly explaining how to calculate the dispersion measure of FRBs.
We are very grateful to the anonymous referee for many insightful comments.
TG acknowledges the support by the Ministry of Science and Technology of Taiwan through grant 105-2112-M-007-003-MY3.




\bibliographystyle{mnras}
\bibliography{FRB_mnras} 




\bsp	
\label{lastpage}
\end{document}